\documentstyle[12pt]{article}

\begin{document}
\setcounter{page}{1} \pagestyle{plain} \vspace{1cm}
\begin{center}
\Large{\bf Minimal length, maximal momentum and thermodynamics of black body radiation}\\
\small \vspace{1cm}
 {\bf H. Shababi\footnote{hedie.shababi@gmail.com}}\\
\vspace{0.5cm} {\it Department of Physics, Islamic Azad
University,\\
Sari Branch, Sari, Iran}
\end{center}
\vspace{1.5cm}
\begin{abstract}
In this paper we study thermodynamics of black body radiation in the
presence of quantum gravitational effects through a Generalized
Uncertainty Principle (GUP) that admits both a minimal measurable
length and a maximal momentum. We focus on quantum gravity induced
modification of thermodynamical quantities in this framework. Some
important issues such as the generalized Planck distribution,
Generalized Wien's law and Generalized Dulong-Petit law are studied
with details.
\end{abstract}
\vspace{1.5cm}
\newpage

\section{Introduction}
Various approaches to quantum gravity such as String Theory and
Doubly Special Relativity predict the existence of a minimal
measurable length or a maximum observable momentum . These theories
argue that near the Planck scale, the Heisenberg Uncertainty
Principle should be replaced by the so called \emph{Generalized
Uncertainty Principle}( GUP) (see [1-5] and references therein).
Modification of the standard uncertainty relation of ordinary
quantum mechanics by incorporation of quantum gravity effects
requires  a reformulation of several important physical laws such as
thermodynamical laws of black body radiation. The importance of
black body radiation lies in the fact that cosmic microwave
background radiation (CMB) is shown to have the same spectrum as the
black body radiation. Since CMB physics today has obtained a very
appreciable place in modern cosmology, any quantum gravitational
correction imposed on this issue will provide a better framework to
understand the real world. On the other hand, these type of study
may provide direct clue to test quantum gravity ideas in the lab.
Planck's early analysis of black body radiation was based on
statistical arguments of black-body radiation within a cavity. When
we consider quantum gravitational modifications,  some important
laws that related to the black body radiation should be corrected.
In this framework, the spectral energy density of black body
radiation gets modified and this modification causes modifications
to other thermodynamical quantities.  Although these modifications
are important only in very high energy scales, but understanding
phenomenological aspects of these results will shed light both in
the formulation of the ultimate quantum gravity proposal and
possible detection of these quantum gravity effects in experiments.
Thermodynamics of black body radiation in the presence of minimal
measurable length has been studied by some authors [6,7]. However,
these studies are not complete since they ignore the fact that a
minimal measurable length essentially requires the existence of a
maximal momentum encoded in the duality of position-momentum spaces
or uncertainty principle. In fact, based on Doubly Special
Relativity theories a test particle's momentum has upper limit of
the order of Planck momentum [8-10]. Existence of a maximal
measurable momentum for a test particle modifies the results of the
mentioned studies considerably. This is because the existence of a
natural cutoff on momentum restricts the number of physically
accessible modes considerably. The purpose of this paper is mainly
to address this issue. We introduce a GUP that admits both a minimal
length and a maximal momentum and we reformulate black body
radiation in this framework. Some important issues such as
generalization of Planck distribution, equipartition theorem,
Stefan-Boltzmann law, Wien's law and Dulong-Petit law will be
studied with details and some important thermodynamical quantities
such as the entropy and specific heat of black body radiation are
calculated in the presence of quantum gravity effects. We also
compare our results with those results that are obtained by ignoring
the role of maximal momentum. For simplicity, in which follows we
set $\hbar=k_{\beta}=c=1$.

\section{Quantum Gravity Effects and Black Body Radiation}
\subsection{GUP with minimal length and maximal momentum}
In the context of the Doubly Special Relativity (DSR)(see [11] for
review), one can show that a test particle's momentum cannot be
arbitrarily imprecise. In fact, there is an upper bound for momentum
fluctuations [8,9]. As a nontrivial assumption, this may lead to a
maximal measurable momentum for a test particle [10]. In this
framework, the GUP that predicts both a minimal observable length
and a maximal momentum can be written as follows [10]
\begin{equation}
\Delta x\Delta p\geq\frac{1}{2}\bigg[1+\Big(\frac{\alpha
L_{p}}{\sqrt{\langle p^2\rangle}}+4\alpha^2L_{p}^{2}\Big)(\Delta
p)^2+4\alpha^2L_{p}^{2}\langle p\rangle^2-2\alpha L_{p}\sqrt{\langle
p^2\rangle}\bigg].
\end{equation}
Since $(\Delta p)^2=\langle p^2\rangle-\langle p\rangle^2$,  by
setting $\langle p\rangle=0$ for simplicity, we find
\begin{equation}
\Delta x\Delta p\geq \frac{1}{2}\bigg[1-2\alpha L_{p}(\Delta
p)+4\alpha^2L_{p}^{2}(\Delta p)^2\bigg].
\end{equation}
It is easy to show how this setup leads to a maximal momentum. To
show this end, we note that the absolute minimal measurable length
in our setup is given by $\Delta x_{min}(\langle
p\rangle=0)\equiv\Delta x_0=\alpha L_{p} $. Due to duality of
position and momentum operators, it is reasonable to assume $\Delta
x_{min}\propto\Delta p_{max}$.  Now, saturating the inequality in
relation (2), we find
\begin{equation}
2(\Delta x\Delta p)=\bigg(1-2\alpha L_{p}(\Delta p)+4\alpha^2
L_{p}^{2}(\Delta p)^2\bigg).
\end{equation}
This results in
\begin{equation}
(\Delta p)^2-\frac{(2\Delta x+2\alpha
L_{p})}{4\alpha^2L_{p}^{2}}(\Delta
p)+\frac{1}{4\alpha^2L_{p}^{2}}=0.
\end{equation}
So, we find
\begin{equation}
(\Delta p_{max})^2-\frac{(2\Delta x_{min}+2\alpha
L_{p})}{4\alpha^2L_{p}^{2}}(\Delta
p_{max})+\frac{1}{4\alpha^2L_{p}^{2}}=0.
\end{equation}
Now using the value of $\Delta x_{min}$, we find
\begin{equation}
(\Delta p_{max})^2-\frac{1}{\alpha L_{p}}(\Delta
p_{max})+\frac{1}{4\alpha^2L_{p}^{2}}=0.
\end{equation}
The solution of this equation is
\begin{equation}
\Delta p_{max} =\frac{1}{2\alpha L_{p}}.
\end{equation}
So, there is an upper bound on particle's momentum uncertainty. As a
nontrivial assumption, we assume that this maximal uncertainty in
particle's momentum is indeed the maximal measurable momentum. This
is of the order of Planck momentum. We note that neglecting a factor
of $\frac{1}{2}$ for simplicity in our forthcoming arguments, the
GUP formulated as (2) gives the following generalized commutation
relation
\begin{equation}
[x,p] = i\Big(1 - \alpha L_{p} p + 2\alpha^2 L_{p}^{2}p^2\Big).
\end{equation}
This is equivalent to set $p\longrightarrow p\Big(1 - \alpha L_{p} p
+ 2\alpha^2 L_{p}^{2} p^2\Big)$. Note that the term $-\alpha L_{p}
p$ that was absent in previous analysis of the black body radiation
[6,7], is related to the existence of maximal momentum and provides
the basic difference of our analysis with previous works.

In the presence of a minimal measurable length and maximal
particle's momentum as natural cutoffs, the spectrum of black body
radiation should be modified. Because of these modifications, the de
Broglie relation is modified too
\begin{equation}
\lambda\simeq \frac{1}{p}(1-\alpha L_{p}p+2\alpha^{2} L_{p} ^2 p^2)
\end{equation}
and therefore,
\begin{equation}
\ E \simeq \nu (1-\alpha L_{p} \nu+2\alpha^{2} L_{p}^2 \nu^2)\,,
\end{equation}
where we have supposed $ \Delta p\simeq p$ and $\Delta
x\simeq \lambda$.\\
Now we consider photons in a cubic box with the length of $L$ and
volume of $V=L^3 $. According to the boundary condition, the
photons' wavelengths are equal to $\frac{1}{\lambda}=\frac{n}{2 L}$,
where $n$ is a positive integer. By considering the above
conditions, we assume that the de Broglie relation is left
unchanged. Therefore the photons have momenta given by
$$p=\frac{n}{2L}.$$ So the momentum space is divided into cells of
volume $V_{p}=(\frac{1}{2L})^3 = \frac{1}{8V}$. Now it follows that
the number of modes with momentum in the interval $[p,\,p+dp]$ is
given by $$g(p) dp = 8 \pi V p^2 dp.$$  Similarly for oscillators in
a box the number of modes in an infinitesimal frequency
interval$[\nu,\nu+d\nu]$  would be written by the following standard
formula
\begin{equation}
\ g(\nu) d\nu = 8 \pi V \nu^2 d\nu
\end{equation}\\
According to (10), the average energy of each oscillator would be
given by
\begin{equation}
\bar{E} = \bigg(\frac{\nu}{e^\frac{\nu}{T}-1}\bigg) \bigg[1-\alpha
L_{p} \nu
\Big(1-\frac{\frac{\nu}{T}}{1-e^\frac{-\nu}{T}}\Big)+2\alpha^{2}
L_{p}^2\nu^2 \Big(
1-\frac{\frac{\nu}{T}}{1-e^\frac{-\nu}{T}}\Big)\bigg]
\end{equation}
This is the generalization of the equipartition theorem in the
presence of minimal length and maximal momentum. We see that in the
limit of $\ L _ {p}\rightarrow0$\ we find the ordinary quantum mechanics result.\\
Now we want to find the modified energy density of the black body
radiation at temperature $T$ and frequency interval $[\nu , \nu+d\nu
]$. In general the energy density is given by
\begin{equation}
\ u_{\nu} (T)d\nu = \frac{\bar{E} g(\nu)d\nu}{V}\,.
\end{equation}
Now according to the modifications induced by the existence of
minimal length and maximal momentum, the energy density appears in
the following form
\begin{equation}
\ u_{\nu} (T)d\nu = 8\pi \Big(\frac{\nu ^ 3 d\nu}{e ^ \frac{\nu}{T}
-1}\Big)\bigg[1-\alpha L_{p} \nu \Big(
1-\frac{\frac{\nu}{T}}{1-e^\frac{-\nu}{T}}\Big)+2\alpha^{2}
L_{p}^2\nu^2 \Big(
1-\frac{\frac{\nu}{T}}{1-e^\frac{-\nu}{T}}\Big)\bigg]
\end{equation}
This is actually the generalized Planck distribution for black body
radiation in the presence of the quantum gravity effects encoded in
the GUP with minimal length and maximal momentum. In our forthcoming
arguments we use this relation as our primary input. \\

\subsection{Energy density}
By integrating of Eq. (14) on frequency, we can calculate the energy
density of black body. This integral gives

$$u(T)= \frac{8}{15} \pi^5 T^4 +\frac{640}{63} \pi^7 T^6 \alpha^{2}
L_{p}^2+128 \pi T^2 \alpha L_{p} m_{p}^3
Li_{2}(e^\frac{m_{p}}{T})+32 \pi T \alpha L_{p} m_{p}^4
\ln(1-e^\frac{m_{p}}{T})$$
$$
-384 \pi T^3\alpha L_{p} m_{p}^2 Li_{3}(e^\frac{m_{p}}{T})+768 \pi
T^4 \alpha L_{p} m_{p} Li_{4}(e^\frac{m_{p}}{T})-80 \pi T \alpha^{2}
L_{p}^2 m_{p}^5 \ln(1-e^\frac{m_{p}}{T})$$
$$
-400 \pi T^2 \alpha^{2} L_{p}^2 m_{p}^4 Li_{2}(e^\frac{m_{p}}{T})
+1600\pi T^3 \alpha^{2} L_{p}^2 m_{p}^3
Li_{3}(e^\frac{m_{p}}{T})-4800 \pi T^4 \alpha^{2} L_{p}^2 m_{p}^2
Li_{4}(e^\frac{m_{p}}{T})$$
$$
+9600 \pi T^5 \alpha^{2} L_{p}^2 m_{p} Li_{5}(e^\frac{m_{p}}{T}) - 2
\pi m_{p}^4+768 \pi T^5 \alpha L_{p} \xi(5)-768 \pi T^5 \alpha L_{p}
Li_{5}(e^\frac{m_{p}}{T})-9600 \pi T^6 \alpha^{2}$$
 $$
  m_{p} L_{p}^2 Li_{6}(e^\frac{m_{p}}{T})+48 \pi
T^4 Li_{4}(e^\frac{m_{p}}{T}) +8 \pi T  m_{p}^3
\ln(1-e^\frac{m_{p}}{T})+24 \pi T^2 m_{p}^2
Li_{2}(e^\frac{m_{p}}{T})
$$
\begin{equation}
-48 \pi T^3 m_{p} Li_{3}(e^\frac{m_{p}}{T})-\frac{224}{35}\pi
m_{p}^5 \alpha L_{p}
\frac{e^\frac{m_{p}}{T}}{e^\frac{m_{p}}{T}-1}+\Big(\frac{840}{63}
e^\frac{m_{p}}{T}+\frac{168}{63}\Big)\pi \alpha^{2} m_{p}^6 L_{p}^2
\end{equation}
where $Li_{s}(z)$ is the Polylogarithm function defined as
$$Li_{s}(z)=\Sigma_{k=1}^{\infty}\frac{z^{k}}{k^{s}}$$
This equation shows the modified Stefan-Boltzmann law and $ m_{p}$
is the Planck mass. We can use the expansion of the Polylogarithm
function to obtain the following result

$$u(T)= \frac{8}{15} \pi^5 T^4 +\frac{640}{63} \pi^7 T^6 \alpha^{2}
L_{p}^2+\pi T^2 \alpha L_{p} m_{p}^3 \big(128 e^\frac{m_{p}}{T}+32
e^\frac{2m_{p}}{T}\big)+32 \pi T \alpha L_{p} m_{p}^4
\ln(1-e^\frac{m_{p}}{T})$$
$$
- \pi T^3\alpha L_{p} m_{p}^2 \big(384 e^\frac{m_{p}}{T}+48
e^\frac{2m_{p}}{T}\big)+\pi T^4 \alpha L_{p} m_{p}\big(768
e^\frac{m_{p}}{T}+48 e^\frac{2m_{p}}{T}\big) -80 \pi T \alpha^{2}
L_{p}^2 m_{p}^5 \ln(1-e^\frac{m_{p}}{T})$$
$$
- \pi T^2 \alpha^{2} L_{p}^2 m_{p}^4 \big(400 e^\frac{m_{p}}{T}+100
e^\frac{2m_{p}}{T}\big) +\pi T^3 \alpha^{2} L_{p}^2 m_{p}^3
\big(1600 e^\frac{m_{p}}{T}+200 e^\frac{2m_{p}}{T}\big)- \pi T^4
\alpha^{2} L_{p}^2 m_{p}^2 \big(4800 e^\frac{m_{p}}{T}+$$
 $$300
e^\frac{2m_{p}}{T}\big) + \pi T^5 \alpha^{2} L_{p}^2 m_{p} \big(9600
e^\frac{m_{p}}{T}+300 e^\frac{2m_{p}}{T}\big) - 2 \pi m_{p}^4+768
\pi T^5 \alpha L_{p} \xi(5)- \pi T^5 \alpha L_{p} \big(768
e^\frac{m_{p}}{T}+24 e^\frac{2m_{p}}{T}\big)$$
 $$
 - \pi T^6
\alpha^{2} m_{p} L_{p}^2 \big(9600
e^\frac{m_{p}}{T}+150e^\frac{2m_{p}}{T}\big)+ \pi T^4 \big(48
e^\frac{m_{p}}{T}+3 e^\frac{2m_{p}}{T}\big) +8 \pi T  m_{p}^3
\ln(1-e^\frac{m_{p}}{T})+ \pi T^2 m_{p}^2 \big(24 e^\frac{m_{p}}{T}+
$$
\begin{equation}
6 e^\frac{2m_{p}}{T}\big)- \pi T^3 m_{p} \big(48 e^\frac{m_{p}}{T}+6
e^\frac{2m_{p}}{T}\big)- \frac{224}{35}\pi m_{p}^5 \alpha L_{p}
\frac{e^\frac{m_{p}}{T}}{e^\frac{m_{p}}{T}-1}+ \Big(\frac{840}{63}
e^\frac{m_{p}}{T}+\frac{168}{63}\Big)\pi \alpha^{2} m_{p}^6 L_{p}^2
\end{equation}
It should be mentioned that we considered up to second order of
correction terms in each expansion.

\subsection{Modified Wien's law}

In ordinary quantum mechanics we have the following relation between
the wavelength at which the energy density distribution maximizes
and corresponding temperature of black body radiation
\begin{equation}
\lambda_{max}=\frac{C}{T}\,,
\end{equation}
where C is Wien's constant. In quantum gravity era this relation
should be modified. Since $u_{\lambda}(T)=\nu^2 u_{\nu}(T)$ and
$\nu=\frac{1}{\lambda}$, we rewrite Eq. (14) in terms of wavelength
to find
\begin{equation}
u_{\lambda}(T)=\frac{8\pi}{\lambda^5\Big(e^\frac{1}{\lambda
T}-1\Big)}-\frac{8\pi \alpha L_{p}}{\lambda^6\Big(e^\frac{1}{\lambda
T}-1\Big)}\bigg(1-\frac{\frac{1}{\lambda T}}{1-e^\frac{-1}{\lambda
T}}\bigg)+\frac{16\pi \alpha^{2}
L_{p}^2}{\lambda^7\Big(e^\frac{1}{\lambda
T}-1\Big)}\bigg(1-\frac{\frac{1}{\lambda T}}{1-e^\frac{-1}{\lambda
T}}\bigg)
\end{equation}
Now for finding modified Wien's law we should calculate the extremum
value of Eq. (18). In our calculations we use the approximation
$e^\frac{1}{\lambda T}=1+\frac{1}{\lambda T}$. So we have the
following relation as the generalized Wien's law in our GUP
framework
$$
\lambda=\frac{1}{12 T}+\frac{\bigg(-18 \alpha L_{p}T+432 \alpha^2
L_{p}^2 T^2+1+6\sqrt{3}\alpha L_{p}T\sqrt{-128 \alpha L_{p} T +7
+1728 \alpha^2  L_{p}^2 T^2}\bigg)^\frac{1}{3}}{12 T}$$
\begin{equation}
-\frac{12 \alpha L_{p} T-1}{12 T\bigg(-18 \alpha L_{p} T+432\alpha^2
L_{p}^2 T^2+1+6\sqrt{3}\alpha L_{p}T \sqrt{-128 \alpha L_{p} T
+7+1728\alpha^2 L_{p}^2 T^2 }\bigg)^\frac{1}{3}}
\end{equation}\\
The first term on the right hand side is the standard Wien's law.
The other terms are corrections imposed from quantum gravity
considerations. We note that there are also two imaginary values for
$\lambda $ that are not acceptable on physical ground. According to
this formula we see that when we consider minimal length and maximal
momentum, the standard Wien's law attains some correction terms that
are temperature dependent. These corrections are important only in
quantum gravity regime. It is important to note that due to quantum
gravitational effects, the $\lambda_{max}$ will attains a small
shift and this shift itself is temperature dependent. This small
wavelength shift can be attributed to the nature of spacetime
manifold at the Planck scale [5].

\section{Some other thermodynamical properties}

Now we can use modified energy density (Eq.(16)) to derive specific
heat capacity and the entropy of black body radiation. So we need
the total energy of the system which is defined as
\begin{equation}
U(T)= V u(T)
\end{equation}
and is given by the following relation

$$U(T)= \frac{8}{15} V\pi^5 T^4 +\frac{640}{63}V \pi^7 T^6 \alpha^{2}
L_{p}^2+\pi T^2 V\alpha L_{p} m_{p}^3 \big(128 e^\frac{m_{p}}{T}+32
e^\frac{2m_{p}}{T}\big)+32 V\pi T \alpha L_{p} m_{p}^4
\ln(1-e^\frac{m_{p}}{T})$$
$$
- \pi T^3 V\alpha L_{p} m_{p}^2 \big(384 e^\frac{m_{p}}{T}+48
e^\frac{2m_{p}}{T}\big)+\pi T^4 V\alpha L_{p} m_{p}\big(768
e^\frac{m_{p}}{T}+48 e^\frac{2m_{p}}{T}\big) -80 \pi V T \alpha^{2}
L_{p}^2 m_{p}^5 \ln(1-e^\frac{m_{p}}{T})$$
$$
- \pi T^2 V\alpha^{2} L_{p}^2 m_{p}^4 \big(400 e^\frac{m_{p}}{T}+100
e^\frac{2m_{p}}{T}\big) +\pi T^3 V\alpha^{2} L_{p}^2 m_{p}^3
\big(1600 e^\frac{m_{p}}{T}+200 e^\frac{2m_{p}}{T}\big)- \pi V T^4
\alpha^{2} L_{p}^2 m_{p}^2 \big(4800 e^\frac{m_{p}}{T}+$$
 $$300
e^\frac{2m_{p}}{T}\big) + \pi T^5 V \alpha^{2} L_{p}^2 m_{p}
\big(9600 e^\frac{m_{p}}{T}+300 e^\frac{2m_{p}}{T}\big) - 2 V \pi
m_{p}^4+768 \pi V T^5 \alpha L_{p} \zeta(5)- \pi V T^5 \alpha L_{p}
\big(768 e^\frac{m_{p}}{T}+$$
$$24 e^\frac{2m_{p}}{T}\big)
 - \pi T^6 V
\alpha^{2} m_{p} L_{p}^2 \big(9600 e^\frac{m_{p}}{T}+150
e^\frac{2m_{p}}{T}\big)+ \pi V T^4 \big(48 e^\frac{m_{p}}{T}+3
e^\frac{2m_{p}}{T}\big) +8 \pi V T  m_{p}^3
\ln(1-e^\frac{m_{p}}{T})+
$$
$$
\pi V T^2 m_{p}^2 \big(24 e^\frac{m_{p}}{T}+6
e^\frac{2m_{p}}{T}\big)- \pi V T^3 m_{p} \big(48 e^\frac{m_{p}}{T}+6
e^\frac{2m_{p}}{T}\big)- \frac{224}{35}\pi V m_{p}^5 \alpha L_{p}
\frac{e^\frac{m_{p}}{T}}{e^\frac{m_{p}}{T}-1}+ \Big(\frac{840}{63}
e^\frac{m_{p}}{T}+\frac{168}{63}\Big)\times$$
\begin{equation}
\pi V \alpha^{2} m_{p}^6
L_{p}^2\quad\quad\quad\quad\quad\quad\quad\quad\quad\quad\quad\quad\quad\quad\quad\quad\quad\quad\quad\quad\quad\quad\quad\quad\quad\quad\quad\quad\quad\quad\quad\quad\quad\
\end{equation}

The specific heat is defined as
\begin{equation}
C_{V}=\Big(\frac{\partial U}{\partial T}\Big)_{V=cte}=T
\Big(\frac{\partial S}{\partial T}\Big)_{V=cte}\,.
\end{equation}\\
So, we have
\begin{equation}
\frac{\partial S}{\partial T}=\frac{C_{V}}{T}
 \end{equation}\\

Therefore the specific heat capacity of black body radiation in the
presence of a minimal length and a maximal momentum is given by

$$ C_{V}=\frac{32}{15} \pi^5 V T^3 +\frac{1280}{21} \pi^7 V T^5
\alpha^2 L_{p}^2+\frac{32 \pi V m_{p}^5 \alpha L_{p}
e^\frac{m_{p}}{T}}{T\big(1-e^\frac{m_{p}}{T}\big)}-\frac{80 V \pi
L_{p}^2 \alpha^2 m_{p}^6
e^\frac{m_{p}}{T}}{T\big(1-e^\frac{m_{p}}{T}\big)}$$
$$ +2\pi T V \alpha L_{p} m_{p}^3\big(128 e^\frac{m_{p}}{T}+32
e^\frac{2m_{p}}{T}\big)-3\pi T^2 V \alpha L_{p} m_{p}^2\big(384
e^\frac{m_{p}}{T}+48e^\frac{2m_{p}}{T}\big)
$$
$$
+4\pi T^3 V \alpha L_{p} m_{p}\big(768 e^\frac{m_{p}}{T}+48
e^\frac{2m_{p}}{T}\big)-2\pi T V \alpha^2 L_{p}^2 m_{p}^4\big(400
e^\frac{m_{p}}{T}+100 e^\frac{2m_{p}}{T}\big) $$
$$ +3\pi T^2 V \alpha^2
L_{p}^2 m_{p}^3\big(1600 e^\frac{m_{p}}{T}+200
e^\frac{2m_{p}}{T}\big)-4\pi T^3 V \alpha^2 L_{p}^2 m_{p}^2\big(4800
e^\frac{m_{p}}{T}+300 e^\frac{2m_{p}}{T}\big)
$$
$$
+5\pi T^4 V \alpha^2 L_{p}^2 m_{p}\big(9600 e^\frac{m_{p}}{T}+300
e^\frac{2m_{p}}{T}\big)-6\pi T^5 V \alpha^2 L_{p}^2 m_{p}\big(9600
e^\frac{m_{p}}{T}+150 e^\frac{2m_{p}}{T}\big)
$$
$$+\frac{8\pi V
m_{p}^4 e^\frac{m_{p}}{T}}{T\big(1-e^\frac{m_{p}}{T}\big)}+\frac{32
\pi V m_{p}^6 \alpha L_{p} e^\frac{m_{p}}{T}}{ 5
T^2\big(e^\frac{m_{p}}{T}-1\big)}-\frac{32 \pi V m_{p}^6 \alpha
L_{p} e^\frac{2m_{p}}{T}}{5
T^2\big(e^\frac{m_{p}}{T}-1\big)^2}+3840\pi T^4 V \alpha L_{p}
\zeta(5)$$
$$
+\pi T^3 V \alpha L_{p} m_{p}\big(768 e^\frac{m_{p}}{T}+48
e^\frac{2m_{p}}{T}\big)+2\pi T V  m_{p}^2\big(24 e^\frac{m_{p}}{T}+6
e^\frac{2m_{p}}{T}\big)-3\pi V T^2 m_{p}\big(48 e^\frac{m_{p}}{T}$$
$$
+6 e^\frac{2m_{p}}{T}\big)+32 \pi\alpha V L_{p} m_{p}^4
Ln\big(1-e^\frac{m_{p}}{T})+\pi\alpha V L_{p} m_{p}^4 \big(-128
e^\frac{m_{p}}{T}-64 e^\frac{2m_{p}}{T}\big)$$
$$ +\pi T V \alpha L_{p} m_{p}^3\big(384
e^\frac{m_{p}}{T}+96 e^\frac{2m_{p}}{T}\big)+\pi T^2 V \alpha L_{p}
m_{p}^2\big(-768 e^\frac{m_{p}}{T}-96 e^\frac{2m_{p}}{T}\big)$$
$$
-80 \pi V\alpha^2 L_{p}^2 m_{p}^5 Ln\big(1-e^\frac{m_{p}}{T})+\pi V
\alpha^2 L_{p}^2 m_{p}^5\big(400 e^\frac{m_{p}}{T}+200
e^\frac{2m_{p}}{T}\big)-\pi T V \alpha^2 L_{p}^2 m_{p}^4$$
$$\big
(1600 e^\frac{m_{p}}{T}+400 e^\frac{2m_{p}}{T}\big)+\pi V
T^2\alpha^2 L_{p}^2 m_{p}^3 \big(4800 e^\frac{m_{p}}{T}+600
e^\frac{2m_{p}}{T}\big)-\pi V T^3\alpha^2 L_{p}^2 m_{p}^2 \big(9600
e^\frac{m_{p}}{T}$$
$$
+600 e^\frac{2m_{p}}{T}\big)-5\pi V T^4\alpha L_{p} \big(768
e^\frac{m_{p}}{T}+24 e^\frac{2m_{p}}{T}\big)+\pi V T^4\alpha^2
L_{p}^2 m_{p}^2 \big(9600 e^\frac{m_{p}}{T}+300
e^\frac{2m_{p}}{T}\big)$$
$$-\frac{40 m_{p}^7 V e^\frac{m_{p}}{T}\pi \alpha^2 L_{p}^2}{3 T^2}- \pi V
m_{p} T^2 \big(48 e^\frac{m_{p}}{T}+6e^\frac{2m_{p}}{T}\big)+4 \pi V
T^3 \big(48 e^\frac{m_{p}}{T}+3e^\frac{2m_{p}}{T}\big)$$
\begin{equation}
 +8 \pi V m_{p}^3
Ln\big(1-e^\frac{m_{p}}{T}\big)-\pi V m_{p}^3 \big(24
e^\frac{m_{p}}{T}+12e^\frac{2m_{p}}{T}\big)+\pi V m_{p}^2 T \big(48
e^\frac{m_{p}}{T}+12e^\frac{2m_{p}}{T}\big)
\end{equation}\\

Now we can use Eq. (23) to obtain entropy of black body radiation.
We have
$$ \frac{C_{V}}{T}=\frac{32}{15} \pi^5 V T^2 +\frac{1280}{21} \pi^7 V
T^4 \alpha^2 L_{p}^2+\frac{32 \pi V m_{p}^5 \alpha L_{p}
e^\frac{m_{p}}{T}}{T^2\big(1-e^\frac{m_{p}}{T}\big)}-\frac{80 V \pi
L_{p}^2 \alpha^2 m_{p}^6
e^\frac{m_{p}}{T}}{T^2\big(1-e^\frac{m_{p}}{T}\big)}$$
$$ +2\pi V \alpha L_{p} m_{p}^3\big(128 e^\frac{m_{p}}{T}+32
e^\frac{2m_{p}}{T}\big)-3\pi T V \alpha L_{p} m_{p}^2\big(384
e^\frac{m_{p}}{T}+48e^\frac{2m_{p}}{T}\big)
$$
$$
+4\pi T^2 V \alpha L_{p} m_{p}\big(768 e^\frac{m_{p}}{T}+48
e^\frac{2m_{p}}{T}\big)-2\pi V \alpha^2 L_{p}^2 m_{p}^4\big(400
e^\frac{m_{p}}{T}+100 e^\frac{2m_{p}}{T}\big) $$
$$ +3\pi T V \alpha^2
L_{p}^2 m_{p}^3\big(1600 e^\frac{m_{p}}{T}+200
e^\frac{2m_{p}}{T}\big)-4\pi T^2 V \alpha^2 L_{p}^2 m_{p}^2\big(4800
e^\frac{m_{p}}{T}+300 e^\frac{2m_{p}}{T}\big)
$$
$$
+5\pi T^3 V \alpha^2 L_{p}^2 m_{p}\big(9600 e^\frac{m_{p}}{T}+300
e^\frac{2m_{p}}{T}\big)-6\pi T^4 V \alpha^2 L_{p}^2 m_{p}\big(9600
e^\frac{m_{p}}{T}+150 e^\frac{2m_{p}}{T}\big)
$$
$$+\frac{8\pi V
m_{p}^4
e^\frac{m_{p}}{T}}{T^2\big(1-e^\frac{m_{p}}{T}\big)}+\frac{32 \pi V
m_{p}^6 \alpha L_{p} e^\frac{m_{p}}{T}}{ 5
T^3\big(e^\frac{m_{p}}{T}-1\big)}-\frac{32 \pi V m_{p}^6 \alpha
L_{p} e^\frac{2m_{p}}{T}}{5
T^3\big(e^\frac{m_{p}}{T}-1\big)^2}+3840\pi T^3 V \alpha L_{p}
\zeta(5)$$
$$
+\pi T^2 V \alpha L_{p} m_{p}\big(768 e^\frac{m_{p}}{T}+48
e^\frac{2m_{p}}{T}\big)+2\pi V  m_{p}^2\big(24 e^\frac{m_{p}}{T}+6
e^\frac{2m_{p}}{T}\big)-3\pi V T m_{p}\big(48 e^\frac{m_{p}}{T}$$
$$
+6 e^\frac{2m_{p}}{T}\big)+\frac{32 \pi\alpha V L_{p} m_{p}^4
Ln\big(1-e^\frac{m_{p}}{T}\big)}{T}+\frac{\pi\alpha V L_{p} m_{p}^4
\big(-128 e^\frac{m_{p}}{T}-64 e^\frac{2m_{p}}{T}\big)}{T}$$
$$ +\pi  V \alpha L_{p} m_{p}^3\big(384
e^\frac{m_{p}}{T}+96 e^\frac{2m_{p}}{T}\big)+\pi T V \alpha L_{p}
m_{p}^2\big(-768 e^\frac{m_{p}}{T}-96 e^\frac{2m_{p}}{T}\big)$$
$$
-\frac{80 \pi V\alpha^2 L_{p}^2 m_{p}^5
Ln\big(1-e^\frac{m_{p}}{T})}{T}+\frac{\pi V \alpha^2 L_{p}^2
m_{p}^5\big(400 e^\frac{m_{p}}{T}+200
e^\frac{2m_{p}}{T}\big)}{T}-\pi V \alpha^2 L_{p}^2 m_{p}^4$$
$$\big
(1600 e^\frac{m_{p}}{T}+400 e^\frac{2m_{p}}{T}\big)+\pi V T\alpha^2
L_{p}^2 m_{p}^3 \big(4800 e^\frac{m_{p}}{T}+600
e^\frac{2m_{p}}{T}\big)-\pi V T^2\alpha^2 L_{P}^2 m_{p}^2 \big(9600
e^\frac{m_{p}}{T}$$
$$
+600 e^\frac{2m_{p}}{T}\big)-5\pi V T^3\alpha L_{p} \big(768
e^\frac{m_{p}}{T}+24 e^\frac{2m_{p}}{T}\big)+\pi V T^3\alpha^2
L_{p}^2 m_{p}^2 \big(9600 e^\frac{m_{p}}{T}+300
e^\frac{2m_{p}}{T}\big)$$
$$-\frac{40 m_{p}^7 V e^\frac{m_{p}}{T}\pi \alpha^2 L_{p}^2}{3 T^3}- \pi V
m_{p}T \big(48 e^\frac{m_{p}}{T}+6e^\frac{2m_{p}}{T}\big)+4 \pi V
T^2 \big(48 e^\frac{m_{p}}{T}+3e^\frac{2m_{p}}{T}\big)$$
\begin{equation}
 +\frac{8 \pi V m_{p}^3
Ln\big(1-e^\frac{m_{p}}{T}\big)}{T}-\frac{\pi V m_{p}^3 \big(24
e^\frac{m_{p}}{T}+12e^\frac{2m_{p}}{T}\big)}{T}+\pi V m_{p}^2
\big(48 e^\frac{m_{p}}{T}+12e^\frac{2m_{p}}{T}\big)
\end{equation}\\
Now by integrating Eq. (25) we can find the entropy of black body
radiation as follows
$$
S=\frac{32}{45}\pi^5 V T^3+\frac{256}{21}\pi^7 V \alpha^2 L_{p}^2
T^5 + 960\pi V \alpha L_{p} T^4 \zeta(5)+64 \pi V \alpha L_{p}
m_{p}^3 \bigg[4 m_{p} Ei(1,-\frac{m_{p}}{T})+$$
$$4 e^\frac{m_{p}}{T} T+ 2
m_{p} Ei(1,-\frac{2 m_{p}}{T})+e^\frac{2 m_{p}}{T} T\bigg]-72 \pi V
\alpha L_{p} m_{p}^2 \bigg[8 m_{p}^2 Ei(1,-\frac{m_{p}}{T})+8
e^\frac{m_{p}}{T} T^2+ 8 m_{p}T e^\frac{m_{p}}{T}$$
 $$+4 m_{p}^2
Ei(1,-\frac{2 m_{p}}{T})+e^\frac{2 m_{p}}{T} T^2 +2 T e^\frac{2
m_{p}}{T} m_{p}\bigg]+64 \pi V \alpha L_{p} m_{p} \bigg[8 m_{p}^3
Ei(1,-\frac{m_{p}}{T})+8 e^\frac{m_{p}}{T} T m_{p}^2+ 8 m_{p}T^2
e^\frac{m_{p}}{T}$$
 $$+16 T^3 e^\frac{m_{p}}{T}+4 m_{p}^3
Ei(1,-\frac{2 m_{p}}{T})+2 T m_{p}^2 e^\frac{ 2 m_{p}}{T}+T^2 m_{p}
e^\frac{ 2 m_{p}}{T}+ T^3 e^\frac{ 2 m_{p}}{T} \bigg]-200 \pi V
\alpha^2 L_{p}^2 m_{p}^2 \bigg[4 m_{p} Ei(1,-\frac{m_{p}}{T})$$
$$+4
e^\frac{m_{p}}{T} T+ 2 m_{p} Ei(1,-\frac{2 m_{p}}{T})+e^\frac{2
m_{p}}{T} T\bigg]+300 \pi V \alpha^2 L_{p}^2 m_{p}^3 \bigg[8 m_{p}^2
Ei(1,-\frac{m_{p}}{T})+8 e^\frac{m_{p}}{T} T^2 + 8 m_{p}T
e^\frac{m_{p}}{T}+4 m_{p}^2 $$
$$Ei(1,-\frac{3 m_{p}}{T})+T^2
e^\frac{ 2 m_{p}}{T}+2 T e^\frac{ 2 m_{p}}{T} m_{p}\bigg]-400 \pi V
\alpha^2 L_{p}^2 m_{p}^2 \bigg[8 m_{p}^3 Ei(1,-\frac{m_{p}}{T})+8
e^\frac{m_{p}}{T} T m_{p}^2 + 8 m_{p}T^2 e^\frac{m_{p}}{T}$$
$$+16
T^3 e^\frac{m_{p}}{T}+  4 m_{p}^3 Ei(1,-\frac{2 m_{p}}{T})+ 2 T
e^\frac{ 2 m_{p}}{T} m_{p}^2+ T^2 e^\frac{ 2 m_{p}}{T} m_{p}^2 +T^3
e^\frac{ 2 m_{p}}{T} \bigg]+125 \pi V \alpha^2 L_{p}^2 m_{p}
\bigg[16 m_{p}^4 Ei(1,-\frac{m_{p}}{T})$$
$$+16 e^\frac{m_{p}}{T} T
m_{p}^3 + 16 m_{p}^2 T^2 e^\frac{m_{p}}{T}+32 m_{p} T^3
e^\frac{m_{p}}{T}+96  T^4 e^\frac{m_{p}}{T}+8 m_{p}^4 Ei(1,-\frac{2
m_{p}}{T})+ 4 T e^\frac{ 2 m_{p}}{T} m_{p}^3+2 T^2 e^\frac{ 2
m_{p}}{T} m_{p}^2$$
$$ +2 T^3 e^\frac{ 2 m_{p}}{T} + 3 T^4  e^\frac{
2 m_{p}}{T} \bigg]-30 \pi V \alpha^2 L_{p}^2 m_{p} \bigg[16 m_{p}^5
Ei(1,-\frac{m_{p}}{T})+16 e^\frac{m_{p}}{T} T m_{p}^4 + 16 m_{p}^3
T^2 e^\frac{m_{p}}{T}+32 m_{p}^2 T^3 e^\frac{m_{p}}{T}$$
$$+96 T^4
e^\frac{m_{p}}{T} m_{p}+384 T^5 e^\frac{m_{p}}{T} +8 m_{p}^5
Ei(1,-\frac{2 m_{p}}{T})+ 4 T e^\frac{ 2 m_{p}}{T} m_{p}^4+2 T^2
e^\frac{ 2 m_{p}}{T} m_{p}^3+ 2 T^3 e^\frac{ 2 m_{p}}{T} m_{p}^2+3
T^4 e^\frac{ 2 m_{p}}{T} m_{p}$$
$$+6 T^5 e^\frac{ 2 m_{p}}{T}
\bigg]+16 \pi V \alpha L_{p} m_{p} \bigg[8 m_{p}^3
Ei(1,-\frac{m_{p}}{T})+8 e^\frac{m_{p}}{T} T m_{p}^2 + 8 m_{p} T^2
e^\frac{m_{p}}{T}+16
 T^3 e^\frac{m_{p}}{T}+4 m_{p}^3
Ei(1,-\frac{2 m_{p}}{T})$$
$$+2 e^\frac{2 m_{p}}{T} T m_{p}^2+
e^\frac{2 m_{p}}{T} T^2 m_{p}+T^3 e^\frac{2 m_{p}}{T}\bigg]+12 \pi V
m_{p}^2 \bigg[4 m_{p} Ei(1,-\frac{m_{p}}{T})+4 e^\frac{m_{p}}{T} T
 +2 m_{p} Ei(1,-\frac{2 m_{p}}{T})+T e^\frac{2 m_{p}}{T}\bigg]$$
 $$-9\pi V  m_{p} \bigg[8 m_{p}^2
Ei(1,-\frac{m_{p}}{T})+8 e^\frac{m_{p}}{T} T^2 + 8 m_{p} T
e^\frac{m_{p}}{T}+4 m_{p}^2 Ei(1,-\frac{2 m_{p}}{T})+T^2 e^\frac{2
m_{p}}{T}+2 T e^\frac{2 m_{p}}{T} m_{p}\bigg]$$
$$-64 \pi V \alpha L_{p} m_{p}^4 \bigg[2
Ei(1,-\frac{m_{p}}{T})+Ei(1,-\frac{2 m_{p}}{T})\bigg]+96 \pi V
\alpha L_{p} m_{p}^3 \bigg[4 m_{p} Ei(1,-\frac{m_{p}}{T})+4
e^\frac{m_{p}}{T} T  + $$
$$2 Ei(1,-\frac{2 m_{p}}{T}){m_{p}} +Te^\frac{2 m_{p}}{T}\bigg]-96 \pi V
\alpha L_{p} m_{p}^2 \bigg[8 m_{p} Ei(1,-\frac{m_{p}}{T})+8
e^\frac{m_{p}}{T} T+2 Ei(1,-\frac{2 m_{p}}{T}){m_{p}} +Te^\frac{2
m_{p}}{T}\bigg]$$
$$+200 \pi V \alpha^2 L_{p}^2 m_{p}^5 \bigg[2
Ei(1,-\frac{m_{p}}{T})+Ei(1,-\frac{2 m_{p}}{T})\bigg]-400 \pi V
\alpha^2 L_{p}^2 m_{p}^4 \bigg[4 m_{p} Ei(1,-\frac{m_{p}}{T})+4
e^\frac{m_{p}}{T} T  +$$
$$ 2 Ei(1,-\frac{2 m_{p}}{T})+T e^\frac{2
m_{p}}{T}\bigg]+300 \pi V \alpha^2 L_{p}^2 m_{p}^3 \bigg[8 m_{p}^2
Ei(1,-\frac{m_{p}}{T})+8 e^\frac{m_{p}}{T} T^2+8 e^\frac{m_{p}}{T} T
m_{p}+4 Ei(1,-\frac{2 m_{p}}{T}) m_{p}^2$$
$$ +T^2 e^\frac{2
m_{p}}{T}+2 T e^\frac{2 m_{p}}{T} m_{p}\bigg]-200 \pi V \alpha^2
L_{p}^2 m_{p}^2 \bigg[8 m_{p}^3 Ei(1,-\frac{m_{p}}{T})+8
e^\frac{m_{p}}{T} T m_{p}^2+8 e^\frac{m_{p}}{T} T^2 m_{p}+16 T^3
e^\frac{m_{p}}{T}+$$
$$4 Ei(1,-\frac{2 m_{p}}{T}) m_{p}^3+2 T
e^\frac{2 m_{p}}{T} m_{p}^2+T^2 e^\frac{2 m_{p}}{T} m_{p}+T^3
e^\frac{2 m_{p}}{T}\bigg]-10 \pi V \alpha L_{p} \bigg[16
Ei(1,-\frac{m_{p}}{T})m_{p}^4+16 e^\frac{m_{p}}{T} T m_{p}^3$$
$$+16
e^\frac{m_{p}}{T} T^2 m_{p}^2+32 e^\frac{m_{p}}{T} T^3 m_{p}+96
e^\frac{m_{p}}{T} T^4+8 Ei(1,-\frac{2 m_{p}}{T})m_{p}^4 +4 T
e^\frac{2 m_{p}}{T} m_{p}^3+2 T^2 e^\frac{2 m_{p}}{T} m_{p}^2+2 T^3
e^\frac{2 m_{p}}{T} m_{p}$$
$$+3 T^4 e^\frac{2 m_{p}}{T}\bigg]+25 \pi V \alpha^2 L_{p}^2 m_{p}^2 \bigg[16
Ei(1,-\frac{m_{p}}{T})m_{p}^4+16 e^\frac{m_{p}}{T} T m_{p}^3 +16
e^\frac{m_{p}}{T} T^2 m_{p}^2+32 e^\frac{m_{p}}{T} T^3 m_{p}+96
e^\frac{m_{p}}{T} T^4$$
$$+8 Ei(1,-\frac{2 m_{p}}{T})m_{p}^4 +4 T
e^\frac{2 m_{p}}{T} m_{p}^3+2 T^2 e^\frac{2 m_{p}}{T} m_{p}^2+2 T^3
e^\frac{2 m_{p}}{T} m_{p} +3 T^4 e^\frac{2
m_{p}}{T}\bigg]+\frac{40}{3}m_{p}^6 \pi V \alpha^2 L_{p}^2 e^\frac{
m_{p}}{T} $$
$$-3 \pi V m_{p} \bigg[8
Ei(1,-\frac{m_{p}}{T})m_{p}^2+8 e^\frac{m_{p}}{T} T^2 +8
e^\frac{m_{p}}{T} T m_{p}+4 Ei(1,-\frac{2 m_{p}}{T})m_{p}^2+T^2
e^\frac{2 m_{p}}{T}+2 T e^\frac{2 m_{p}}{T} m_{p}\bigg]$$
$$+4 \pi V  \bigg[8
Ei(1,-\frac{m_{p}}{T})m_{p}^3+8 e^\frac{m_{p}}{T} T m_{p}^2 +8
e^\frac{m_{p}}{T} T^2 m_{p}+16 e^\frac{m_{p}}{T} T^3+4 Ei(1,-\frac{2
m_{p}}{T})m_{p}^3+2 T e^\frac{2 m_{p}}{T} m_{p}^2$$
$$+ T^2 e^\frac{2
m_{p}}{T} m_{p}+T^3 e^\frac{2 m_{p}}{T}\bigg]-12 m_{p}^3 \pi V
\bigg[2 Ei(1,-\frac{m_{p}}{T})+Ei(1,-\frac{2 m_{p}}{T})\bigg]+12
m_{p}^2 \pi V \bigg[4 Ei(1,-\frac{m_{p}}{T})m_{p}$$
\begin{equation}
+4 T e^\frac{m_{p}}{T}+2 Ei(1,-\frac{2 m_{p}}{T})+T e^\frac{2
m_{p}}{T}
\bigg]+...\quad\quad\quad\quad\quad\quad\quad\quad\quad\quad\quad\quad\quad\quad\quad\quad\quad\quad\quad\quad\quad\quad\quad\quad
\end{equation}
Where $Ei(a,z)$ is the exponential integral defined
as$$Ei(a,z)=z^{a-1} \Gamma(1-a, z)$$

We know from [5,6] that by considering only a minimal length GUP,
the corrections of entropy and specific heat capacity of black body
radiation contain only the odd power of $T$. But here with a GUP
that admits both minimal length and maximal momentum, we see that
both odd and even powers of $T$ are present in entropy and specific
heat relations. In the standard limit we recover the results of
ordinary quantum mechanics.

Figures $1$, $2$ and $3$ show the variation of energy density,
entropy and specific heat capacity of black body radiation versus
temperature. The departure of quantum gravity results from the
standard results are enhanced in the high temperature limit. We note
that energy density, entropy and specific heat capacity of black
body radiation in the presence of the quantum gravity effects
encoded in GUP (2) are generally larger than corresponding
quantities in ordinary quantum mechanics. So, quantum gravity
corrections on the black body spectrum are generally
temperature-dependent and increase the thermodynamical quantities
relative to their ordinary quantum mechanical values.

\begin{figure}[htp]
\begin{center}\includegraphics{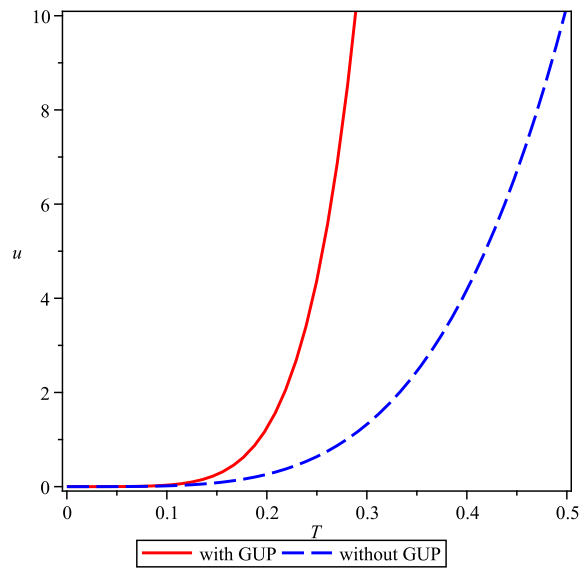} \vspace{8cm}
\end{center}
\caption{\small{Energy density of black body radiation versus its
temperature.}}
\end{figure}

\begin{figure}[htp]
\begin{center}\includegraphics{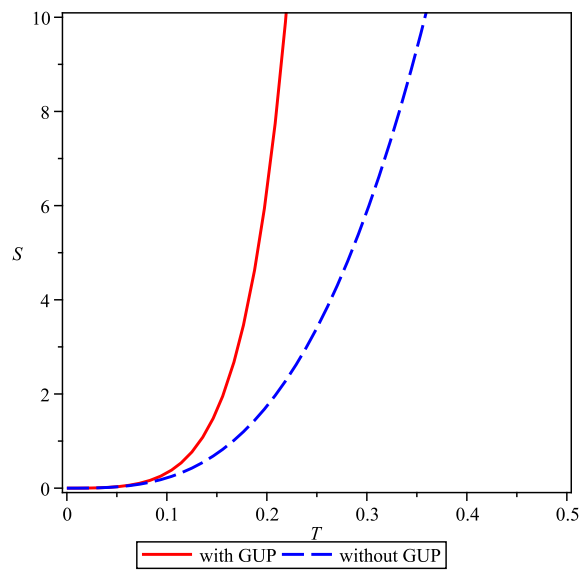} \vspace{7cm}
\end{center}
\caption{\small {entropy of black body radiation versus its
temperature.}}
\end{figure}

\begin{figure}[htp]
\begin{center}\includegraphics{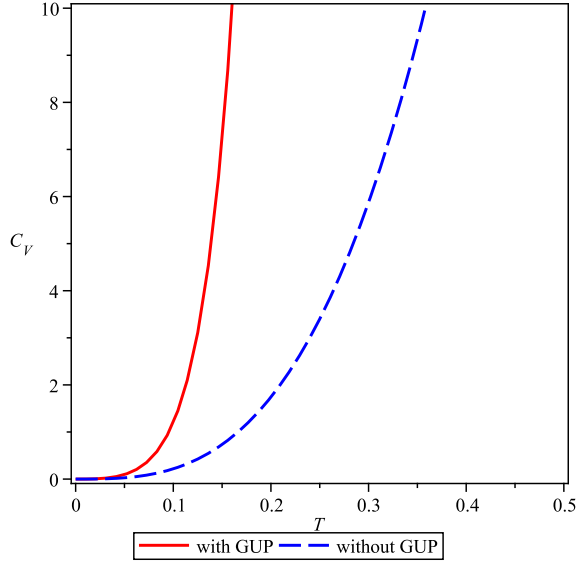} \vspace{7cm}
\end{center}
\caption{\small {Specific Heat Capacity of black body radiation
versus its temperature.}}
\end{figure}
\newpage

\section{Modified Dulong-Petit law}
\subsection{Standard framework}
An statement of the Dulong-Petit law in modern term is that
regardless of the nature of the substance or crystal, the specific
heat capacity $C$ of a solid substance (measured in Joule per Kelvin
per Kilogram) is equal to $\frac{3R}{M}$, where $R$ is the gas
constant (measured in Joule per Kelvin per Mole) and $M$ is the
molar mass (measured in Kilogram per Mole). Thus the heat capacity
per Mole of many solids is $3R$. A system of vibrations in a
crystalline solid can be modeled by considering harmonic oscillator
potentials along each degree of freedom. Then the free energy of
system can be written as
\begin{equation}
F=N \varepsilon_{0}+K_{\beta}T  \sum_{\alpha}
\log\Big(1-e^\frac{-\hbar \omega_{\alpha}}{k_{\beta}T}\Big)\,,
\end{equation}
where the index $\alpha$ sums over all the degrees of freedom. We
consider the case where $K_{\beta}T\gg \hbar \omega_{\alpha}$. So we
have
$$1-e^\frac{-\hbar \omega_{\alpha}}{k_{\beta}T}\approx \frac{\hbar \omega_{\alpha}}{k_{\beta}T}$$
and therefore
\begin{equation}
F=N \varepsilon_{0}+K_{\beta}T \sum_{\alpha} \log\Big(\frac{\hbar
\omega_{\alpha}}{k_{\beta}T})
\end{equation}
Now we define geometric mean frequency by
$$\log\bar{\omega}=\frac{1}{M}\sum_{\alpha}\log \omega_{\alpha}$$
where M measures the total number of degrees of freedom. Thus we
have
\begin{equation}
F=N \varepsilon_{0}- M K_{\beta}T \log k_{\beta}T + M k_{\beta}T
\log \hbar\bar{\omega}
\end{equation}
By using the definition
\begin{equation}
U=F-T \bigg(\frac{\partial F}{\partial T}\bigg)_{V}
\end{equation}
 we have
$$U=N \varepsilon_{0}+M K_{\beta} T$$
This gives specific heat capacity as
\begin{equation}
C_{V}=\bigg(\frac{\partial U}{\partial T}\bigg)_{V}=M K_{\beta}
\end{equation}
which is independent of the temperature .\\

\subsection{GUP framework}
Now we want to rewrite the above equations in the presence of
minimal length and maximal momentum effects encoded in GUP (2). In
this situations
\begin{equation}
F=N \varepsilon_{0}+K_{\beta}T \sum \log\Big(1-e^\frac{-E(1-\beta
E+2\beta E^2)}{k_{\beta}T}\Big)
\end{equation}
and we have
$$ K_{\beta}T\gg E(1-\beta E+2\beta E^2).$$
So we can use the approximation
$$ 1-e^\frac{-E(1-\beta E+2 \beta E^2)}{K_{\beta}T} \approx
\frac{E(1-\beta E+2\beta E^2)}{K_{\beta}T}$$ Therefore
\begin{equation}
F=N \varepsilon_{0}+K_{\beta}T\sum\log\bigg(\frac{E(1-\beta E
+2\beta E^2)}{K_{\beta}T}\bigg)
\end{equation}
and
\begin{equation}
\log\bar{E}=\frac{1}{M}\sum \log E(1-\beta E+2\beta E^2)
\end{equation}
Now by using of Eq. (12), we have find
$$F=N\varepsilon_{0}+M K_{\beta}T
\log\bigg(\frac{\nu}{e^\frac{\nu}{T}-1}\Big(1-\alpha L_{p} \nu
(1-\frac{\frac{\nu}{T}}{1-e^\frac{-\nu}{T}})$$
\begin{equation}
 +2\alpha^2 L_{p}^2
\nu^2(1-\frac{\frac{\nu}{T}}{1-e^\frac{-\nu}{T}})\Big)\bigg)-M
K_{\beta} T \log K_{\beta}T
\end{equation}
To calculate specific heat capacity we should calculate total energy
from Eq.(30) . We find
$$
U=N\varepsilon_{0}+M K_{\beta}T -\frac{M
K_{\beta}T^2}{e^\frac{\nu}{T}-1}\bigg[\frac{1}{1-\alpha L_{p}
\nu\big(1-\frac{\nu}{T(1-e^\frac{-\nu}{T})}\big)+2\alpha^2 L_{p}^2
\nu^2\big(1-\frac{\nu}{T(1-e^\frac{-\nu}{T})}\big)} \bigg]\times$$
$$
\bigg[-\alpha L_{p} \nu
\big[\frac{\nu}{T^2(1-e^\frac{-\nu}{T})}-\frac{\nu^2
e^\frac{-\nu}{T}}{T^3(1-e^\frac{-\nu}{T})^2}\big]+ 2\alpha^2 L_{p}^2
\nu^2 \big[\frac{\nu}{T^2(1-e^\frac{-\nu}{T})}-\frac{\nu^2
e^\frac{-\nu}{T}}{T^3(1-e^\frac{-\nu}{T})^2}\big]\bigg]$$
\begin{equation}
-\frac {M K_{\beta}\nu^2\bigg[1-\alpha L_{p}
\nu\big(1-\frac{\nu}{T(1-e^\frac{-\nu}{T})}\big) +2\alpha^2 L_{p}^2
\nu^2\big(1-\frac{\nu}{T(1-e^\frac{-\nu}{T})}\big)\bigg]e^\frac{-\nu}{T}}
{e^\frac{\nu}{T}-1}\quad\quad\quad\quad\quad\quad
\end{equation}
Now after finding $U$, we can calculate the specific heat capacity
to find
$$C_{V}=M K_{\beta}-T \Bigg[\frac{1}{\nu\bigg(1-\alpha
L_{p}\nu\big(1-\frac{\nu}{T(1-e^\frac{-\nu}{T})}\big)+2\alpha^2
L_{p}^2
\nu^2\big(1-\frac{\nu}{T(1-e^\frac{-\nu}{T})}\big)\bigg)}\bigg[2 M
K_{\beta}\big[\frac{1}{e^\frac{\nu}{T}-1}[\nu[$$
 $$-\alpha
L_{p}\nu[\frac{\nu}{T^2(1-e^\frac{-\nu}{T})}-\frac{\nu^2
e^\frac{-\nu}{T}}{T^3(1-e^\frac{-\nu}{T})^2}]+2\alpha^2
L_{p}^2\nu^2[\frac{\nu}{T^2(1-e^\frac{-\nu}{T})}-\frac{\nu^2
e^\frac{-\nu}{T}}{T^3(1-e^\frac{-\nu}{T})^2}]]]$$
$$+\frac{\nu^2\bigg(1-\alpha
L_{p}\nu\big(1-\frac{\nu}{T(1-e^\frac{-\nu}{T})}\big)+2\alpha^2
L_{p}^2 \nu^2\big(1-\frac{\nu}{T(1-e^\frac{-\nu}{T})}\big)\bigg)
e^\frac{\nu}{T}}{(e^\frac{\nu}{T}-1) T^2}\big]\bigg]$$
$$+\frac{1}{\nu\bigg(1-\alpha
L_{p}\nu\big(1-\frac{\nu}{T(1-e^\frac{-\nu}{T})}\big)+2\alpha^2
L_{p}^2
\nu^2\big(1-\frac{\nu}{T(1-e^\frac{-\nu}{T})}\big)\bigg)}\bigg[ M K_
{\beta}T\big[\frac{1}{e^\frac{\nu}{T}-1}[\nu[-\alpha L_{p}\nu[$$
$$-\frac{2\nu}{T^3(1-e^\frac{-\nu}{T})}+\frac{4\nu^2
e^\frac{-\nu}{T}}{T^4(1-e^\frac{-\nu}{T})^2}-\frac{\nu^3
e^\frac{-\nu}{T}}{T^5(1-e^\frac{-\nu}{T})^2}-\frac{2\nu^3
(e^\frac{-\nu}{T})^2}{T^5(1-e^\frac{-\nu}{T})^3}]+2\alpha^2 L_{p}^2
\nu^2[-\frac{2\nu}{T^3(1-e^\frac{-\nu}{T})}+\frac{4\nu^2
e^\frac{-\nu}{T}}{T^4(1-e^\frac{-\nu}{T})^2}$$
$$-\frac{\nu^3
e^\frac{-\nu}{T}}{T^5(1-e^\frac{-\nu}{T})^2}-\frac{2\nu^3
(e^\frac{-\nu}{T})^2}{T^5(1-e^\frac{-\nu}{T})^3}]]]+\frac{2\nu^2
\Big [-\alpha L_{p} \nu
\big[\frac{\nu}{T^2(1-e^\frac{-\nu}{T})}-\frac{\nu^2
e^\frac{-\nu}{T}}{T^3(1-e^\frac{-\nu}{T})^2}\big]}{(e^\frac{\nu}{T}-1)^2
T^2}$$
$$+\frac{2\alpha^2 L_{p}^2
\nu^2 \big[\frac{\nu}{T^2(1-e^\frac{-\nu}{T})}-\frac{\nu^2
e^\frac{-\nu}{T}}{T^3(1-e^\frac{-\nu}{T})^2}\big]\big](e^\frac{\nu}{T})}{(e^\frac{\nu}{T}-1)^2
T^2}-\frac{\nu^3\bigg(1-\alpha
L_{p}\nu\big(1-\frac{\nu}{T(1-e^\frac{-\nu}{T})}\big)+2\alpha^2
L_{p}^2 \nu^2\big(1-\frac{\nu}{T(1-e^\frac{-\nu}{T})}\big)\bigg)
e^\frac{\nu}{T}}{(e^\frac{\nu}{T}-1)^2 T^4}$$
$$+\frac{2\nu^3\bigg(1-\alpha
L_{p}\nu\big(1-\frac{\nu}{T(1-e^\frac{-\nu}{T})}\big)+2\alpha^2
L_{p}^2 \nu^2\big(1-\frac{\nu}{T(1-e^\frac{-\nu}{T})}\big)\bigg)
(e^\frac{\nu}{T})^2}{(e^\frac{\nu}{T}-1)^3 T^4}-$$
$$\frac{2\nu^2\bigg(1-\alpha
L_{p}\nu\big(1-\frac{\nu}{T(1-e^\frac{-\nu}{T})}\big)+2\alpha^2
L_{p}^2 \nu^2\big(1-\frac{\nu}{T(1-e^\frac{-\nu}{T})}\big)\bigg)
e^\frac{\nu}{T}}{(e^\frac{\nu}{T}-1) T^3}\big]\bigg]$$
$$-\frac{1}{T\bigg(1-\alpha
L_{p}\nu\big(1-\frac{\nu}{T(1-e^\frac{-\nu}{T})}\big)+2\alpha^2
L_{p}^2
\nu^2\big(1-\frac{\nu}{T(1-e^\frac{-\nu}{T})}\big)\bigg)}\bigg[ M K_
{\beta}\big[\frac{1}{e^\frac{\nu}{T}-1}[\nu[-\alpha
L_{p}\nu[\frac{\nu}{T^2(1-e^\frac{-\nu}{T})}$$
$$
-\frac{\nu^2 e^\frac{-\nu}{T}}{T^3(1-e^\frac{-\nu}{T})^2}]+2
\alpha^2
L_{p}^2\nu^2[\frac{\nu}{T^2(1-e^\frac{-\nu}{T})}-\frac{\nu^2
e^\frac{-\nu}{T}}{T^3(1-e^\frac{-\nu}{T})^2}]]]$$
$$
+\frac{\nu^2\bigg(1-\alpha
L_{p}\nu\big(1-\frac{\nu}{T(1-e^\frac{-\nu}{T})}\big)+2\alpha^2
L_{p}^2 \nu^2\big(1-\frac{\nu}{T(1-e^\frac{-\nu}{T})}\big)\bigg)
e^\frac{\nu}{T}}{(e^\frac{\nu}{T}-1)^2
T^2}\bigg]e^\frac{\nu}{T}\bigg]$$
$$
-\frac{1}{\nu\bigg(1-\alpha
L_{p}\nu\big(1-\frac{\nu}{T(1-e^\frac{-\nu}{T})}\big)+2\alpha^2
L_{p}^2
\nu^2\big(1-\frac{\nu}{T(1-e^\frac{-\nu}{T})}\big)\bigg)^2}\bigg[ M
K_ {\beta}T\big[\frac{1}{e^\frac{\nu}{T}-1}[\nu[-\alpha L_{p}\nu[$$
$$
\frac{\nu}{T^2(1-e^\frac{-\nu}{T})}-\frac{\nu^2
e^\frac{-\nu}{T}}{T^3(1-e^\frac{-\nu}{T})^2}]+2\alpha^2 L_{p}^2
\nu^2[\frac{\nu}{T^2(1-e^\frac{-\nu}{T})}-\frac{\nu^2
e^\frac{-\nu}{T}}{T^3(1-e^\frac{-\nu}{T})^2}]]]$$
$$+\frac{\nu^2\bigg(1-\alpha
L_{p}\nu\big(1-\frac{\nu}{T(1-e^\frac{-\nu}{T})}\big)+2\alpha^2
L_{p}^2 \nu^2\big(1-\frac{\nu}{T(1-e^\frac{-\nu}{T})}\big)\bigg)
e^\frac{\nu}{T}}{(e^\frac{\nu}{T}-1) T^2}\bigg]\big[-\alpha
L_{p}\nu[$$
\begin{equation}
\frac{\nu}{T^2(1-e^\frac{-\nu}{T})}-\frac{\nu^2
e^\frac{-\nu}{T}}{T^3(1-e^\frac{-\nu}{T})^2}]+2\alpha^2
L_{p}^2\nu^2[\frac{\nu}{T^2(1-e^\frac{-\nu}{T})}-\frac{\nu^2
e^\frac{-\nu}{T}}{T^3(1-e^\frac{-\nu}{T})^2}]]\bigg]\Bigg]
\end{equation}
In the absence of quantum gravity this relation reduces to the
standard Dulong-Petit law. Note that the GUP-corrected terms are
temperature-dependent as a generic feature of quantum gravity
effects.

\newpage

\section{Summary and Conclusion}
In this paper we have studied the effects of minimal length and
maximal momentum, as natural cutoffs encoded in a generalized
uncertainty principle, on the thermodynamics of black body
radiation. The importance of the black body radiation lies in the
fact that cosmic microwave background radiation is shown to have the
same spectrum as the black body radiation. We have found the
generalization of equipartition theorem, the generalized Planck
distribution, Modified Wien's law and the modified Stefan- Boltzmann
law. In the next step we have found some thermodynamical properties
of black body in the presence of mispecific heat capacity and
entropy. And finally we have calculated the modified Dulong-Petit
law in this framework. We have shown that quantum gravity
corrections are generally temperature-dependent. This is a generic
feature and has its origin probably on the very nature of space-time
at quantum gravity scalesnimal length and maximal momentum such as
the specific heat capacity and entropy. And finally we have
calculated the modified Dulong- Petit law in this framework. We have
shown that quantum gravity corrections are generally
temperature-dependent. This is a generic feature and has its origin
probably on the very nature of space-time at quantum gravity scales.\\

{\bf Acknowledgment}\\
I would like to thank Prof. Kourosh Nozari for insightful comments
and discussion.

\end{document}